\documentstyle[aaspp4]{article}

\lefthead{Taniguchi and Murayama}
\righthead{Dusty Tori in Seyfert Nuclei}

\begin{document}

\title{DUSTY TORI OF SEYFERT NUCLEI 
PROBED BY THE WATER VAPOR MASER EMISSION:
HOW LARGE ARE THE DUSTY TORI ?}

\author{Yoshiaki Taniguchi, \& Takashi Murayama}
\affil{Astronomical Institute, Tohoku University, Aoba, Sendai 980-8578, Japan}

\begin{abstract}

Geometrical and physical properties of dusty tori of Seyfert nuclei probed by the
water vapor maser emission at 22 GHz are discussed.
We assume that the dusty torus has a simple cylindrical form and the maser emission
can be detected only when we observe the torus from almost edge-on views.
The observed low frequency of occurrence of the water vapor
maser emission (less than 10 percent) suggests that the torus is a vertically thin
cylinder whose outer radius between a few pc and $\sim$ 10 pc.
However, the observed masing regions are concentrated in the inner 1 pc 
regions of the torus. This property can be explained by that only the inner 
a few pc regions have physical conditions enough to cause the maser emission;
the temperature is as high as several hundred  K and the density is as high as 
$\sim 10^{10}$ cm$^{-3}$.

\end{abstract}


\section{INTRODUCTION}

The current unified models of active galactic nuclei (AGN) have introduced a dusty
torus which surrounds the central engine. Since the torus is considered to
be optically very thick, the visibility of the central engine is significantly
affected by viewing angles toward AGN (Antonucci \& Miller 1985;
Krolik \& Begelman 1988;  see for a review Antonucci 1993).
The dusty tori emit their energy mostly in mid- and far-infrared wavelengths.
Therefore, infrared spectral energy distributions of AGN have been often utilized to
study physical properties of the dusty tori [Pier \& Krolik 1992b, 1993 
(hereafter PK92 and PK93, respectively); 
Heckman, Chambers, \& Postman 1992; Heckman 1995; Granato, Danese, \& Franceschini 1997;
Taniguchi et al. 1997; Murayama, Mouri, \& Taniguchi 1998].
However, geometrical properties of the dusty tori have not yet been studied directly
because  the dusty tori are too compact to be resolved spatially at mid- and 
far-infrared wavelengths.

On the other hand,
for these several years, we have learned that the water vapor maser emission at 22 GHz
can be used to probe dusty tori directly (Nakai, Inoue, \& Miyoshi 1993; 
Miyoshi et al. 1995; Greenhill et al. 1995a, 1995b, 1996; Gallimore et al. 1996;
Greenhill \& Gwinn 1997).
In particular, the recent VLBA or VLA measurements of the H$_2$O maser emission
of the nearby AGNs, NGC 1068 (Gallimore et al. 1996; Greenhill et al. 1996;
Greenhill \& Gwinn 1997),
NGC 4258 (Miyoshi et al. 1995; Greenhill et al. 1995a, 1995b),
and NGC 4945 (Greenhill, Moran, \& Herrnstein 1997a), have shown that the masing clouds
are located at distances of $\sim$ 0.1 - 1 pc from the nuclei.
Since these distances are almost comparable to those of the dusty tori,
it is suggested that the masing clouds reside in the tori themselves
(e.g., Greenhill et al. 1996).
Therefore, the H$_2$O maser emission provides a useful tool to study 
physical properties of dusty tori as well as dynamical ones
(e.g., Murayama \& Taniguchi 1997 and references therein). 
The recent H$_2$O maser searches [Braatz, Wilson, \& Henkel 1996, 1997
(hereafter BWH97); Greenhill et al. 1997b] also
provides important information.  Although the microphysics of
the maser emission has been investigated in detail 
(Krolik \& Lepp 1989; Neufeld, Maloney, \& Conger 1994), 
the H$_2$O maser data have not yet
fully utilized to study geometrical properties of the dusty tori.
Therefore, in this {\it Letter},
incorporating all the above observational results on the H$_2$O maser emission,
we discuss the geometrical properties of the dusty tori.

\section{DUSTY TORI OF SEYFERT NUCLEI PROBED BY THE WATER VAPOR MASER EMISSION}

\subsection{\it Dusty Torus Models}

Several dusty torus models for AGN have been proposed up to now
(Efstathiou \& Rowan-Robinson 1990; PK92, PK93; Granato \& Danese 1994;
Granato, Danese, \& Franceschini 1997).
Among them, models of PK92 are quite simple and thus are very useful
for investigations of statistical properties of dusty tori (PK93).
Therefore, we adopt the simple torus models of 
PK92 who investigated the thermally reradiated infrared spectra of the compact
dusty tori surrounding the central engine of AGNs
by using a two-dimensional radiative transfer
algorithm. The torus is a cylinder of dust with a uniform density,
characterized by the inner radius ($a$), the outer radius ($b$),
and the full height ($h$).
A half opening angle of the
torus is thus given  as $\theta_{\rm open}$ =  tan$^{-1}$($2a/h$).
The important parameters describing models
are; 1) the inner aspect ratio, ($a/h$), which is coupled with
the covering factor, $f = 1 -
\Omega/4\pi$, where $\Omega$ is the solid angle subtended at
the source not covered by dust,
2) the effective temperature of the torus, $T_{\rm eff}$, which is roughly equal to the
hottest dust temperature in the torus, and  3) the radial and the vertical
Thomson optical depth,
$\tau_r = n_{\rm H} \sigma_{\rm T} (b-a)$ and $\tau_z =  n_{\rm H} \sigma_{\rm T} h$,
where $n_{\rm H}$ is the hydrogen number density and $\sigma_{\rm T}$ is the Thomson
cross section.

Comparing the observed infrared spectra with the model spectra,
PK93 showed that the tori of Seyfert nuclei have $T_{\rm eff} \simeq$ 600 - 
800 K and $a/h \simeq$ 0.3. Such geometrically thick tori may be supported by
the radiation pressure from the central engines (Pier \& Krolik 1992a). 
PK93 also made a case study for a dusty torus of 
the archetypical type 2 Seyfert nucleus of NGC 1068 in detail and 
found that a model with $T_{\rm eff} = 800$ K, 
$a/h$ = 0.3, and $\tau_r = \tau_z = 1$, can explain the observed 
spectral energy distribution of  NGC 1068.
The inner radius of the torus can be given by
$a = 0.59 (T_{\rm eff} / 800~ {\rm K})^{-2} ~ (L_{\rm bol} / 2.4 \times 10^{11} 
L_\odot)^{-0.5}$ pc where $L_{\rm bol}$ is the bolometric luminosity of the 
central engine of NGC 1068. The most reliable estimate of $L_{\rm bol}$ 
of NGC 1068 is given by Pier et al. (1994);
$L_{\rm bol} = 2.2 \times 10^{11} ~ (f_{\rm refl} / 0.01)^{-1} ~
(D / 22~ {\rm Mpc})^2$ where $f_{\rm refl}$ is 
the fraction of nuclear flux reflected into our line of sight
and $D$ is the distance to NGC 1068.
Thus we obtain $a \simeq$ 0.56 pc which is just comparable to the observed
inner radius of the H$_2$O masing region (Greenhill et al. 1996).
The gas mass of the torus of NGC 1068 can be estimated as
$M_{\rm gas} \simeq 560 ~  (T_{\rm eff} / 800~ {\rm K})^{-4} ~
\tau_r ~ (a/h)^{-1} ~ [(\tau_r / \tau_z) (a/h)^{-1} + 2] ~ L_{44} ~ M_\odot ~
\simeq 4.9 \times 10^4 M_\odot$ where $L_{44}$ is the bolometric luminosity of the 
central engine in units of $10^{44}$ erg s$^{-1}$ (PK92).

One remaining problem is {\it how to estimate the outer radius of the torus}.
Since most of the energy in the torus is radiated
through the top to bottom surfaces within $r < a + h$, it is difficult
to constrain $b$ using the infrared spectral energy distribution.
For example, if we adopt a dusty torus with
$a$ = 1 pc, $b$ = 10 pc, $h$ = 2 pc, and $M_{\rm gas} = 10^5 M_\odot$
(e.g., Murayama, Taniguchi, \& Iwasawa 1998), 
this torus gives an average hydrogen number density, $\overline{n}_{\rm H} \simeq 6.6 \times
10^3$ cm$^{-3}$ and  an HI column density for an edge-on view toward the torus,
$N_{\rm HI} =  \overline{n}_{\rm H} (b-a) 
\simeq 1.8 \times 10^{23}$ $(M_{\rm gas}/10^5 M_\odot) ~ (b/ {\rm 10~ pc})^{-1}
(h/ {\rm 2~ pc})^{-1}$ cm$^{-2}$.
The inner aspect ratio of this torus, $a/h=0.5$, gives  a semi-opening angle
of the torus, $\theta_{\rm open} = 45^\circ$, being consistent with
the typical half opening angle of narrow-line regions, $\sim 30^\circ$ - 45$^\circ$
(Pogge 1989; Wilson \&  Tsvetanov 1994;
Schmitt \& Kinney 1996; see also Osterbrock \& Shaw 1988;
Miller \& Goodrich 1990; Lawrence 1991).
Further, hard X-ray observations have shown that 
typical Seyfert 2 galaxies have the HI column 
density of of the order of $10^{23}$ cm$^{-2}$ (e.g., Awaki et al. 1997).
Here it is remembered that the H$_2$O maser emission tends to be detected
in AGNs with higher HI column densities, e.g., $N_{\rm HI} > 10^{23}$ cm$^{-2}$
(BWH97). It is thus suggested that the H$_2$O maser emission can be observed
only when we see the dusty torus from an almost edge-on view. 
If we adopt this hypothesis, we are able to estimate a typical outer radius
statistically using the observed frequency of occurrence of the maser emission
in AGNs.

\subsection{\it A Statistical Size of the Dusty Tori inferred from the 
Frequency of Occurrence of  H$_2$O Maser}

First we give a brief summary of the recent comprehensive survey 
of the H$_2$O maser emission for $\sim$ 350 AGNs by BWH97.
(1) The  detection rate of the H$_2$O maser emission is $\sim$ 5\% at most.
(2) No H$_2$O maser emission has been detected in type 1 Seyferts (hereafter S1s).
(3) Therefore, the H$_2$O maser emission can be seen only in type 2 Seyferts
(hereafter S2s) and LINERs\footnote{LINER = Low Ionization Nuclear
Emission-line Region (Heckman 1980).}. The detection rate of the H$_2$O maser
emission is still low, $\sim$ 7\%, even only for the S2s.
And, (4) the AGN with the H$_2$O maser emission tends to have a higher HI column density,
$N_{\rm HI} \sim 10^{23-24}$ cm$^{-2}$, inferred from the hard X-ray
observations.
In Table 1, the observed detection rates of the H$_2$O maser emission
are summarized for some AGN samples studied by BWH97.
The observations show that the detection rates of the H$_2$O maser emission
are $\simeq$ 5\% among the observed Seyfert nuclei for both the distance-limited
and the magnitude-limited samples.

The important observational properties are both that the H$_2$O maser emission 
has not yet been observed in S1s and  that
the S2s with the H$_2$O maser emission have the higher HI column densities
toward the central engine (BWH97). It is hence suggested strongly
that the maser emission can be detected only when the
dusty torus is viewed from almost edge-on views.
This is advocated by the ubiquitous presence of so-called the main
maser component whose velocity is close to the systemic one
whenever the maser emission is observed because 
this component arises from dense molecular gas clouds
along the line of sight between the background amplifier (the central engine) and us
(e.g., Miyoshi et al. 1995; Greenhill et al. 1995a).

Since the high HI column density is achieved only when we see the torus
within the aspect angle, $\phi =$ tan$^{-1} (h/2b)$
(see Figure 1), we are able to estimate $b$ because the detection rate of
H$_2$O maser, $P_{\rm maser}$, emission can be related to the aspect angle as,
$P_{\rm maser} = N_{\rm maser}/(N_{\rm maser} + N_{\rm non-maser}) 
= {\rm cos}(90^\circ - \phi)$ where 
$N_{\rm maser}$ and $N_{\rm non-maser}$ are the numbers of AGN with
the H$_2$O maser emission and without the H$_2$O maser emission, respectively.
This relation gives $b = h~ [2 {\rm tan}(90^\circ - {\rm cos}^{-1} P_{\rm maser})]^{-1}$.
Table 1 shows that a typical detection rate is $P_{\rm maser} \simeq$ 0.05.
However, this value should be regarded as a lower limit because
some special properties of may be necessary to cause the maser emission
(Wilson 1998). If we take account of new detections of H$_2$O maser emission
from NGC 5793 (Hagiwara et al. 1997) and NGC 3735 (Greenhill et al. 1997b)
which were discovered by two other maser surveys independent from BWH97,
the detection rate may be as high as $\simeq$ 0.1 (Wilson 1998).
Therefore, we estimate $b$ values for the two cases; 1) $P_{\rm maser}$ = 0.05,
and $P_{\rm maser}$ = 0.1.  These two rates correspond to the aspect angles,
$\phi \simeq 2.^\circ9$ and $\phi \simeq 5.^\circ7$, respectively.
In Table 2, we give the estimates of $b$ for three cases, 
$a$ = 0.1, 0.5, and 1 pc.
If $a >$ 1 pc, the HI column density becomes lower than
$10^{23}$ cm$^{-2}$ given $M_{\rm gas} = 10^5 M_\odot$. 
Therefore, it is suggested that the inner radius may be in a range 
between 0.1 pc and 0.5 pc for typical Seyfert nuclei.
The inner radii of the H$_2$O masing regions in NGC 1068, NGC 4258, and NGC
4945 are indeed in this range (Greenhill et al. 1996; Miyoshi et al. 1995;
Greenhill et al. 1997a).
We thus obtain possible sizes of the dusty tori;
($a, b, h$) = (0.1 - 0.5 pc, 1.67 - 8.35 pc, 0.33 - 1.67 pc) 
for $\phi \simeq 5.^\circ7$, and 
($a, b, h$) = (0.1 - 0.5 pc, 3.29 - 16.5 pc, 0.33 - 1.67 pc) 
for $\phi \simeq 2.^\circ9$.
All the cases can achieve $N_{\rm HI} > 10^{23}$ cm$^{-2}$,
being consistent with the observations (BWH97). 

We have shown that the most probable outer radii of the dusty tori are from
a few pc to 10 pc.
On the other hand, the observed outer radii of the H$_2$O masing
regions are between 0.25 pc (NGC 4258: Miyoshi et al. 1995)
and 1 pc (NGC 1068: Greenhill et al. 1996).
In Table 3, we give a summary of the observations.
We can regard that the inner radius of the H$_2$O maser is almost
equal to the inner radius of the torus in each case.
However, it is worth noting that the outer radius of the torus
is much larger than that of the H$_2$O masing region.
Another important aspect is that the masing regions can be observed
in an area of $r < r_{\rm hot}$ for all the objects.
This implies that the H$_2$O maser emission can
arise only from the inner hot (several hundred  K), dense
($\sim 10^{10}$ cm$^{-3}$)  region of the torus (cf. Elitur 1992).
The thermal equilibrium cannot be attained in the torus
and thus the temperature of H$_2$O molecules may not necessarily be
the same as that of the dust.
However, it is unlikely that the molecule temperature
is quite different from that of dust.
In other words, we may conclude that the H$_2$O maser emission can be
generated only in dense, hot molecular clouds in the inner parts of
the dusty tori.

In this {\it Letter}, we assumed for simplicity 
that sufficient optical depth for maser emission occurs 
only if the torus is seen edge-on.
However, we should remind that the same effect could be produced
if tori exhibit a range of intrinsic (e.g. vertical) column densities.
This point shall also be taken into account in future investigations.

\subsection{A Case for Warped Dusty Tori}

Finally we mention about a case for warped dusty tori.
Even if the dusty torus is warped significantly like that of NGC 1068
(Begelman \& Bland-Hawthorn 1997), the HI column density is still high,
$N_{\rm HI} \simeq (l/b) \overline{n}_{\rm H} (b-a) \simeq
 8 \times 10^{22} (M_{\rm gas}/10^5 M_\odot)
~ (b/ {\rm 10~ pc})^{-2} ~ ({\rm sin} \xi / 0.5)^{-1}$ cm$^{-2}$ where
$l$ is the path length of the line of sight within the torus and
$\xi$ is the intersecting angle between the torus plane and the line of sight
(see Figure 2).  In this estimate, we adopt $\xi = 30^\circ$ as a fiducial value.
If the masing condition could be achieved in any parts
of the torus, the detection rate of the H$_2$O maser emission would be higher
significantly than the observed one.
For example, if the aspect angle were $\phi \simeq 30^\circ$,
the detection rate of the H$_2$O maser emission would be
$P_{\rm maser} \sim 0.5/2 \sim 0.25$, being much higher than the
observed rates (see Table 1).
We can reconcile this dilemma if we take account of the inadequate
physical conditions of the outer parts of the tori; i.e.,
the gas density may be too low (e.g., $\overline{n}_{\rm H} \sim 10^4$
cm$^{-3}$) to cause the maser emission and the temperature should be also
too low (e.g., $\sim$ 50 K: PK92).

\acknowledgments

We would like to thank Naomasa Nakai and Toru Yamada for useful discussion
and  the anonymous referee for very useful comments.
TM is a JSPS Research Fellow.
This work was financially supported in part by a Grant-in-Aid for
Scientific Research (No. 07044054) from
the Japanese Ministry of Education, Science, Sports, and Culture.


\clearpage

\begin{table*}
\begin{center}
\begin{tabular}{cccccc}
\hline \hline
Sample & $N_{\rm maser}$ & $N_{\rm total}$ & $P_{\rm maser}$ (\%) \\
\hline
Distance-limited & & & \\
\hline
All (S1+S2+L) & 15 & 278 & 5.4 \\
Seyfert (S1+S2) & 10 & 198 & 5.1 \\
S2 & 10 & 141 & 7.1 \\
\hline
Magnitude-limited & & & \\
\hline
All (S1+S2+L) & 13 & 241 & 5.4 \\
Seyfert (S1+S2) & 8 & 166 & 4.8 \\
S2 & 8 & 112 & 7.1 \\
\hline \hline
\end{tabular}
\end{center}
\tablenum{1}
\caption{A summary of the detection rates of the H$_2$O maser in active galactic nuclei
studied by Braatz, Wilson, \& Henkel (1997) for the various samples}
\end{table*}


\clearpage

\begin{table*}
\begin{center}
\begin{tabular}{ccccccc}
\hline \hline
 & & & \multicolumn{2}{c}{$P_{\rm maser}=0.05$} & \multicolumn{2}{c}{$P_{\rm maser}=0.1$} \\
 & & & \multicolumn{2}{c}{$\phi=2.^\circ9$} & \multicolumn{2}{c}{$\phi=5.^\circ7$} \\
$a$ (pc) & $h$ (pc) &  $r_{\rm hot}$\tablenotemark{a} (pc) & $b$ (pc) & 
$N_{\rm HI}$ (cm$^{-2}$) & $b$ (pc) & $N_{\rm HI}$ (cm$^{-2}$) \\
\hline
0.1 & 0.33 & 0.43 & 3.29 & $3.3 \times 10^{24}$ &1.67 & $6.5\times 10^{24}$ \\
0.5 & 1.67 & 2.17 & 16.5 & $1.3\times 10^{23}$ & 8.35 & $2.6\times 10^{23}$ \\
  1 & 3.30 & 4.30 & 32.9 & $3.3\times 10^{22}$ & 16.7 & $6.5\times 10^{22}$ \\
\hline \hline
\end{tabular}
\end{center}
\tablenotetext{a}{The radius of the hot part in the torus; $r_{\rm hot} = a + h$.}
\tablenum{2}
\caption{Geometrical properties of the dusty tori inferred from the statistics of 
the H$_2$O maser emission}
\end{table*}


\clearpage

\begin{table*}
\begin{center}
\begin{tabular}{cccccccc}
\hline \hline
Galaxy & Type & $D$ & $a=r_{\rm in}$(H$_2$O) & $h$\tablenotemark{a} & 
 $r_{\rm out}$(H$_2$O) & $r_{\rm hot}$ & $b$\tablenotemark{e}  \\
  &  & (Mpc) & (pc) & (pc) & (pc) & (pc) & (pc) \\
\hline
NGC 1068 & S2 & 22 & 0.56\tablenotemark{b} & 1.87 & 1.0\tablenotemark{b} & 2.43 & 9.35 \\
NGC 4258 & L & 6.4 & 0.13\tablenotemark{c} & 0.43 & 0.25\tablenotemark{c} & 0.56 & 2.17 \\
NGC 4945 & S2/L & 3.7 & 0.18\tablenotemark{d} & 0.60 & 0.45\tablenotemark{d} & 0.78 &3.01 \\
\hline \hline
\end{tabular}
\end{center}
\tablenotetext{a}{We assume $a/h=0.3$.}
\tablenotetext{b}{Greenhill et al. (1996)}
\tablenotetext{c}{Miyoshi et al. (1995)}
\tablenotetext{d}{Greenhill et al. (1997a)}
\tablenotetext{e}{The outer radius of the torus for the case of $P_{\rm maser}=0.1$.}
\tablenum{3}
\caption{Active galactic nuclei with spatially-resolved H$_2$O masing regions}
\end{table*}




\newpage


\figcaption{A section of the dusty torus adopted in this paper.
\label{fig1}}

\figcaption{A section of the warped dusty torus.
\label{fig2}}



\begin{thebibliography}{}
\bibitem{} Antonucci, R. 1993, ARA\&A, 31, 473 
\bibitem{} Antonucci, R. R. J., Miller, J. S. 1985, ApJ, 297, 621
\bibitem{} Awaki, H., Ueno, S., Koyama, K., Iwasawa, K., \& Kunieda, H.
           1997, Advances in Space Research, 9, 95
\bibitem{} Begelman, M.\ C., Bland-Hawthorn, J.\ 1997, Nature 385, 22
\bibitem{} Braatz, J. A., Wilson, A. S., \& Henkel, C. 1996, ApJS, 106, 51
\bibitem{} Braatz, J. A., Wilson, A. S., \& Henkel, C. 1997, ApJS, 110, 321 (BWH97)
\bibitem{} Efstathiou, A., \& Rowan-Robinson, M. 1990, MNRAS, 245, 275
\bibitem{} Elitur, M. 1992, Astrophysical Masers (Dordrecht, Klewer), Chap. 10
\bibitem{} Gallimore, J. F., Baum, S. A., O'Dea, C. P., Brinks, E.,
           Pedlar, A. 1996, ApJ, 462, 740
\bibitem{} Granato, G. L., \& Danese, L. 1994, MNRAS, 268, 235
\bibitem{} Granato, G., Danese, L., \& Franceschini, A. 1997, ApJ, 486, 147
\bibitem{} Greenhill, L. J., \& Gwinn, C. R. 1997, Ap \& SS, 248, 261
\bibitem{} Greenhill, L. J., Gwinn, C. R., Antonucci, R, \& Barvanis, R.
           1996, ApJ, 472, L21
\bibitem{} Greenhill, L. J., Herrnstein, J. R., Moran, J. M., Menten, K. M.,
           \& Velusamy, T. 1997b, ApJ, 486, L15
\bibitem{} Greenhill, L. J., Jiang, D. R., Moran, J. M., Reid, M. J.,
           Lo, K. Y.,  \& Claussen, M. J. 1995a, ApJ, 440, 619
\bibitem{} Greenhill, L. J., Henkel, C., Becker, R., Wilson, T. L., \&
           Wouterloot, J. G. A. 1995b, A \& A, 304, 21
\bibitem{} Greenhill, L. J., Moran, J. M., \& Herrnstein, J. R. 1997a, ApJ, 481, L23
\bibitem{} Hagiwara, Y., Kohno, K., Kawabe, R., \& Nakai, N. 1997, PASJ, 49, 171
\bibitem{} Heckman, T. M. 1980, A \& A, 87, 152
\bibitem{} Heckman, T. M. 1995, ApJ, 446, 101 
\bibitem{} Heckman, T. M., Chambers, K. C., \& Postman,  M. 1992, ApJ, 391, 39
\bibitem{} Krolik, J. H., \& Begelman, M. C. 1988, ApJ, 329, 702
\bibitem{} Krolik, J. H., \& Lepp. S. 1989, ApJ, 347, 179
\bibitem{} Lawrence, A. 1991, MNRAS, 252, 586
\bibitem{} Miller, J. S., \& Goodrich, R. W. 1990, ApJ, 355, 456 
\bibitem{} Miyoshi, M., Moran, J., Herrnstein, J., Greenhill, L., Nakai, N.,
           Diamond, P., \& Inoue, M. 1995,  Nat, 373, 127
\bibitem{} Murayama, T., Mouri, H., \& Taniguchi, Y. 1998, ApJ, submitted
\bibitem{} Murayama, T., \& Taniguchi, Y. 1997, PASJ, 49, L13
\bibitem{} Murayama, T., Taniguchi, Y., \& Iwasawa, K. 1998, AJ, 115, 460
\bibitem{} Nakai, N.,  Inoue, M.,  \&  Miyoshi, M. 1993, Nature, 361, 45
\bibitem{} Neufeld, D. A., Maloney, P. R., \& Conger, S. 1994, ApJ, 436, L127
\bibitem{} Osterbrock, D. E., \& Shaw, R. 1988, ApJ, 327,  89
\bibitem{} Pier, E. A., Antonucci, R., Hunt, T., Kriss, G., \& Krolik, J. 1994,
           ApJ, 428, 124
\bibitem{} Pier, E. A., \& Krolik, J. H. 1992a, ApJ, 399, L23
\bibitem{} Pier, E. A., \& Krolik, J. H. 1992b, ApJ, 401, 99 (PK92)
\bibitem{} Pier, E. A., \& Krolik, J. H. 1993, ApJ, 418, 673 (PK93)
\bibitem{} Pogge, R. W., 1989, ApJ, 345, 730
\bibitem{} Schmitt, H. R., \& Kinney, A. L. 1996, ApJ, 463, 498
\bibitem{} Taniguchi, Y., Sato, Y., Kawara, K., Murayama, T., \& Mouri, H.
           1997, A \& A, 318, L1
\bibitem{} Wilson, A. S. 1998, Accretion Processes in Astrophysical Systems:
           Some Like it Hot, in press
\bibitem{} Wilson, A. S., \& Tsvetanov, Z. I. 1994, AJ, 107, 1227
\end{thebibliography}
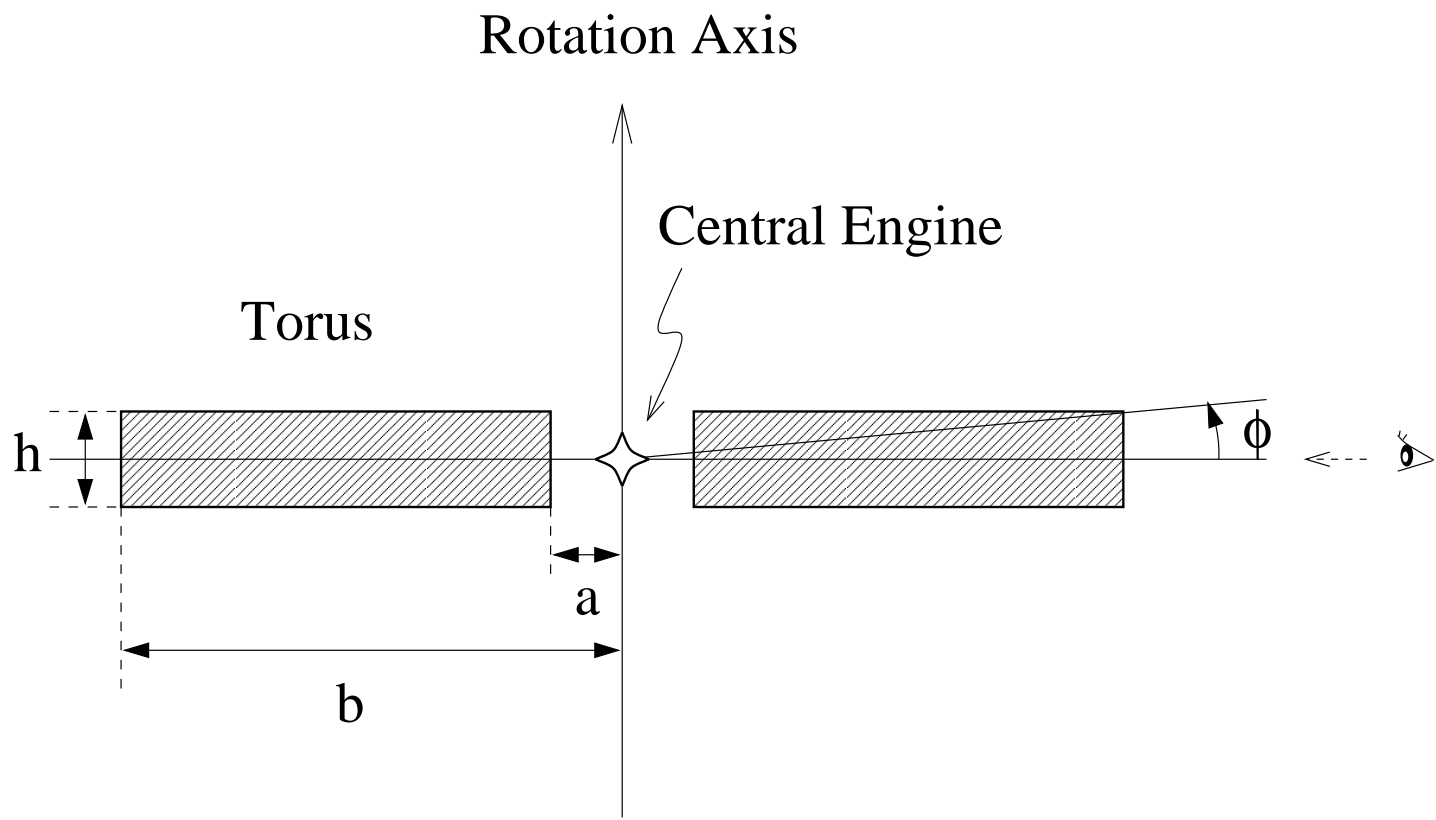
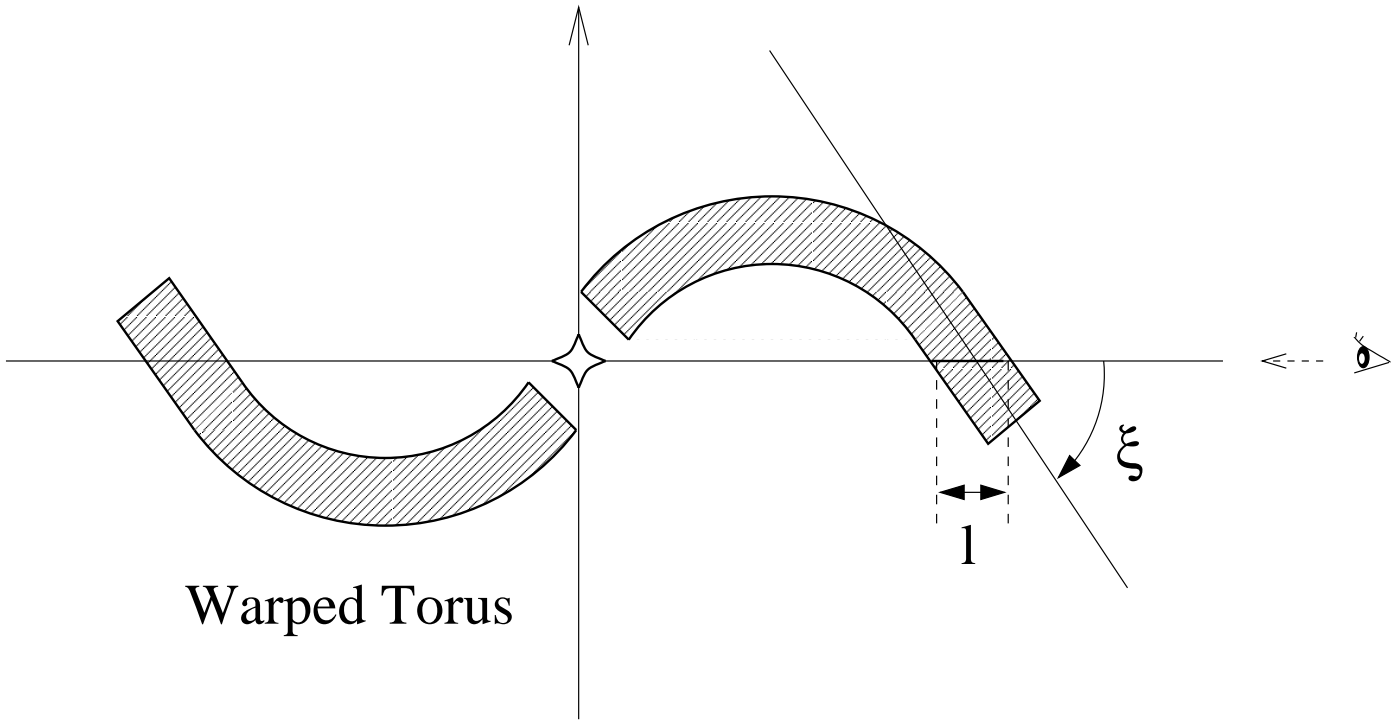
\end{document}